\documentclass{aastex62}

\graphicspath{{./}{figures/}}

\submitjournal{ApJ}

\shorttitle{O I lines pumping}
\shortauthors{Nemer et al.}
\usepackage{amsmath}
\usepackage{color}  
\newcommand{\unit}[1]{\,{\rm #1}}  
\newcommand{\au}{\textsc{au}}
\newcommand{\nH}{n_{\textsc{h}}}

\newcommand{\forbid}[2]{[\textsc{#1}]\,#2\textrm{\AA}} 
\newcommand{\msun}{{\rm M}_\odot}
\newcommand{\lsun}{{\rm L}_\odot}

\begin{document}

\title{The role of FUV Pumping in exciting the [\textsc{Oi}] lines in Protostellar Disks and Winds}

\correspondingauthor{Ahmad Nemer}
\email{anemer@princeton.edu}

\author[0000-0002-0786-7307]{Ahmad Nemer}
\affil{Princeton University \\
4 Ivy Lane \\
Princeton, New Jersey, USA}

\author{Jeremy Goodman}
\affiliation{Princeton University \\
4 Ivy Lane \\
Princeton, New Jersey, USA}

\author{Lile Wang}
\affiliation{Center for Computational Astrophysics \\
Flatiron Institute\\
162 Fifth Avenue\\
New York, USA}



\begin{abstract}

We use Cloudy to re-examine excitation of \forbid{Oi}{6300} and \forbid{Oi}{5577} in the X-ray driven photoevaporative wind models of Owen and collaborators, and in more recent magnetothermal models by Wang et al.  We find that, at the measured accretion luminosities, the FUV radiation would populate the upper levels of the oxygen atom which would eventually contribute to the [\textsc{Oi}] lines. FUV pumping competes with collisions as an excitation mechanism of the [\textsc{Oi}] lines, and they each originate from a distinct region in the protostellar disk environment depending on the specifics of the hydrodynamical model that the simulation was based on. Consequently, the line strengths and shapes of \forbid{Oi}{6300} and \forbid{Oi}{5577} would be affected by the inclusion of FUV pumping in the radiation transport simulations.
\end{abstract}



\section{Introduction} \label{sec:intro}
Planets form in protoplanetary disks around Young Stellar Objects (YSOs). The similarities between the timescales predicted for planet formation and disk dispersal in YSOs suggest that the two processes are coupled. Thus, determining the dominant mechanisms for disk dispersal is essential in understanding the formation of planets. A wide range of mechanisms have been suggested to explain how the disks of young stars lose their mass: viscous accretion(\citep[e.g.][]{Lynden-Bell1974}); photoevaporative disk winds as a result of thermal motion of the gas (\citep{Ercolano2017}); Planet formation \citep{Partnership2015}; MHD winds from small radii (X winds; \citep[e.g.][]{Shu2000}), or from the broader disk radii (\citep[e.g.][]{Konigl-Pudritz2000}). One or more of those mechanisms are responsible for mass loss from protostellar disks, and using emission lines diagnostics presents a way to trace the different dispersal processes. \\

In the presence of magnetic fields and ionized accretion disks, mass ejection in the form of wind is predicted; probably the process is complex and involves multiple sub-processes. A magnetocentrifugal disk wind, due to the residual magnetic field that threads the disk at all distances from the star, an X-wind, which is launched close to the star where the pressure from the stellar magnetic field overcomes the stellar gravity and, a stellar wind, where gas from the central star is expelled from the system (\citealt{Ferreira2013}, and references therein). Under the action of the magnetic field, the gas is accelerated to terminal velocities of a few to hundreds km $s^{-1}$, forming the slow disk winds and the bright jets observed in several young objects \citep[e.g.][]{Frank+2014}. As disk and stellar winds extract angular momentum, they can control accretion \citep[e.g.][]{Turner+2014}.\\

Photoevaporation is another key mass loss process in which gas at extended distances from the disk surface is heated to temperatures such that the particles' thermal energy exceeds its gravitational energy and the gas is able to escape from the system. This effect was first detected in systems where disks are dispersed from around hot stars \citep[e.g.][]{O'dell+Wen1994}. Photoevaporation recently gained more attention due to the detection of high energy photons (FUV, EUV, Xray) from low mass stars that could provide the thermal pressure, through absorption and heating, which would launch a rotating slow wind. Photoevaporation winds are suspected to quickly dissipate the disk, which ends their extended phases of evolution through accretion and outflow (see, e.g., \citealt{Alexander+2014}, and references therein).\\

The presence of a warm, at least partially, ionised disc wind has been confirmed via the observation of a few km/s blue-shift in the profile of the [Ne\textsc{ii}]$12.8\unit{\mu m}$ fine structure line (e.g., \citealt{Gudel+2014} and references therein). Moreover, this line shows two components; a High velocity (HVC) and Low velocity (LVC) components. The HVC of the [Ne\textsc{ii}] is clearly associated with jets/outflows (it has been also spatially resolved toward one source; see \citealt{vanBoekel+2009}), while the profile and peak velocity of the LVC are consistent with a photoevaporative disk wind driven by stellar X-ray or EUV photons \citep{Pascucci+Sterzik2009,Sacco+2012,Baldovin-Saavedra+2012}). The line can be reproduced in EUV-driven-partially-ionized gas models as well as in X-ray-driven models where the gas is largely neutral; these models predict mass loss rates that differ by 2 orders of magnitude \citep[][hereafter EO16]{Ercolano+Owen2016}. \\

More recently, forbidden lines of low-charge and neutral states of heavy elements have provided a potential tool for tracing the slow disk winds. Specifically the \forbid{Oi}{6300} and \forbid{Oi}{5577} lines have been intensively investigated as wind tracers. Like the [Ne\textsc{ii}]~$12.8\unit{\mu m}$ line, the \textsc{Oi} lines are observed with an HVC and LVC component where the HVC likely traces a collimated jet outflow. It is unlikely that an EUV-driven wind could produce the observed luminosities, due to insufficient neutral oxygen in the mostly ionized gas. It is suspected that FUV and/or X-ray radiation could provide the physical conditions for collisional excitation of the [\textsc{Oi}] ($T_e \sim 6000 \unit{K}$ $n_e > 10^6\unit{cm^{-3}}$) without over-ionizing the gas, because of the greater column density (compared to EUV) at which such photons are absorbed. EO16 affirmed that their X-ray photoevaporative models are able to produce the observed luminosities and the line shapes.\\

Furthermore, there has been debate about the excitation mechanism for the \forbid{Oi}{6300} and \forbid{Oi}{5577} lines. \cite{Rigliaco+2013} brought a new perspective to the observation of the \textsc{Oi} LVC. They accurately derived the accretion luminosity for 30 of the \cite{Hartigan+1995} stars from $H_{\alpha}$ emission lines, observed simultaneously with the forbidden lines, which appeared in \cite{Beristain+2001}. They also used high-resolution spectra to distinguish the LVC into a Narrow Component (NC), which probably traces a photoevaporative wind, and a Broad Component (BC) which likely traces an MHD wind in Keplerian motion. They observed a correlation between the accretion luminosity and the \textsc{Oi} lines' luminosity as well as a correlation between the stellar FUV and the lines' luminosity, but there was no correlation with the X-ray luminosity. This led them to the conclusion that the X-ray photoevaporation models are unlikely to be the only mechanism responsible for exciting these lines. They suggest that the origin of the \textsc{Oi} forbidden lines, in addition to photoevaporation, could be due to the dissociation of OH molecules on a surface disk layer that is powered by stellar FUV radiation. \\

Moreover, \cite{Simon+2016} found that both the BC and the NC correlate with the accretion luminosity (which in turn correlates with FUV radiation) over a wide range of accretion luminosities. They also establish a relationship between the FWHM of the BC with the disk inclination and the peak centroid of the line. This brings them to the conclusion that the emission of the BC comes from radii between 0.05 an 0.5 au and that it includes emission from a wind that is broadened by Keplerian rotation. They also observe larger blueshifts for the BC associated with larger accretion luminosities, and with the gravitational binding energy at these radii, the BC is unlikely to trace a photoevaporative wind but rather an MHD one; but the excitation mechanism is yet to be confirmed. The NC, on the other hand, is predicted to be emitted from larger radii (0.5 to 5 au) with Keplerian broadening profiles, but the photoevaporation origin of this line cannot be excluded because of the lower escape energy needed at these radii and the lack of a relationship between centroid velocity and disk inclination. \\

Both the observations of \cite{Rigliaco+2013} and \cite{Natta+2014} invoked inconsistencies between EO10 models and observational data. EO16 have addressed these inconsistencies by modifying their models with the addition of an artificial accretion luminosity, to illuminate the wind but not to drive it, and have handled the effect of neutral hydrogen collisions on the [\textsc{Oi}] lines more carefully. They conclude that their new models are in excellent agreement with the available data, and that the [\textsc{Oi}] lines most likely trace photoevaporative winds. \\

In this paper we conduct radiation-transfer simulations using Cloudy \citep{Ferland+2017} and the density profile based on hydro-dynamical models by Owen et al and Wang et al (unpublished) to understand the effect of the accretion luminosity (more specifically the FUV radiation) and the neutral-hydrogen collisions on the thermal excitation of the \textsc{Oi} forbidden lines. We also want to investigate the dominant physical process responsible for exciting the [\textsc{Oi}] forbidden lines and whether the thermal excitation alone can explain the observed luminosity (as claimed by EO16) or whether there is another physical process, related to the FUV source, that poses a better candidate like OH dissociation (as claimed by \citealt{Rigliaco+2013} and \citealt{Gorti+2011}) or optical pumping of the lines by the FUV photons. Optical pumping in this context means excitation to higher bound levels of the oxygen atom by absorption of FUV photons, followed by a cascade of radiative or collisional de-excitations leading to the upper levels of the forbidden lines. We illustrate the effects of pumping in both EO16/10’s photoevaporative wind and in Wang et al’s (unpublished) magnetothermal wind.  The goal of this article, however, is to demonstrate the effect of pumping on the [\textsc{Oi}] lines, rather than to study whether the lines can distinguish between photoevaporative and magnetothermal winds.  We leave the latter for future work, though we note the recent investigation of the subject by \cite{Weber+2020}.

\section{Results} \label{sec:results}

We present the work we performed, using Cloudy, to simulate the details of the \forbid{Oi}{6300} and 5577\AA{} line emission from a protoplanetary disk environment. The radiative transfer simulation was performed following the work of EO16 in an attempt to reproduce their results. We thank James Owen for supplying the flow field (density, velocity, temperature) underlying Fig 4 of EO16. We represented the radiation source with four blackbody sources with various effective temperatures following the work of EO16. We used a blackbody source with $T_{\rm eff}= 4250 \unit{K}$, $M = 0.7\msun$, and $R_* = 2.5 \unit{R_\odot}$  to represent the photospheric luminosity ($L_*=1.83\lsun$). Another source with $T_{\rm eff}=12000\unit{K}$ was used to simulate the accretion radiation with a luminosity equal to the stellar luminosity (as in the second row of EO16's Table 1). Finally we included two more blackbody sources to mimic the EO16's rather complex X-ray SED: one with $T_{\rm eff} = 10^6 \unit{K}$, and another with $T_{\rm eff} = 10^7\unit{K}$, dividing EO16's X-ray luminosity of $2 \times 10^{30} \unit{ergs/s}$ equally between the two. We use the same elemental abundances as reported by EO16. We also include dust grains (dust-to-gas ratio of $\sim1\%$) with different parameters in our model, and we found that dust (besides the input radiation sources) has the most significant impact on the produced luminosity. Cloudy is capable only of spherical (or slab) geometries.  However, insofar as scattering and optical-depth effects of the O I line are unimportant, we can use Cloudy along radial rays through the (non-spherical) wind, each such ray having its own density profile. The gas starts at the illuminated face of the cloud at $r = 1.74 \times 10^{11} \unit{cm}$ and ends at $r=40\unit{au}$. The density profile that was fed to these Cloudy models was extracted from the results of the Hydrodynamic models computed by Owen et al. The temperature is calculated self-consistently at every zone in Cloudy after solving the statistical equilibrium equations, the ionization balance, and the thermal equilibrium equations. Cloudy includes an extensive network of chemical, collisional and radiative processes to accurately obtain solutions for these equilibrium equations. Cloudy also has the possibility to switch on and off some of those various processes. \\

Below we report the results of the different calculations done to investigate the origin of the \forbid{Oi}{6300} line emission and to reproduce the threshold luminosity $\sim1\times10^{-5}\lsun$ of this line in EO16's models (hereafter the "observed" luminosity), and investigate the main excitation mechanisms. The calculations were performed on a spherical grid with 1000 linearly spaced points in radius between $0.01$ and $40\unit{au}$, and 241 points uniformly spaced in $\sin\theta\in[0,1]$. To get the total luminosity, we integrate the line emissions over radius and over one hemisphere in $(\theta,\phi)$, supposing that the disk obscures the emission from the other hemisphere. To avoid the simulation from crashing, we extrapolate the the density profile from $<0.4\unit{au}$ to the the illuminated face of the cloud, but we exclude emission from radii $<0.4\unit{au}$ because this information is missing from the data supplied by Owen et al. \\

We varied the size and types of grains used in our models to understand the effect they have on the \textsc{Oi} 6300 line emission. We tried using models that don't include grains, different abundances of ISM dust grains (as used by EO16), different types of PAH grains, and 1 micron graphite and silicate grains. We found that some of these models were able to produce the observed luminosity of the \textsc{Oi} 6300 A. In the models that were able to reproduce the observed luminosities for the \textsc{Oi} 6300 and 5577 A lines, the ratio of these lines was closer ($\sim$ 12) to the observed ratio (1 - 10) than the models that did not match (ratio of$\sim$45) the observed luminosities. In our fiducial model Fig. \ref{fig:Cloudy1}(where we use 1 micron dust grains), the emission mainly came from two distinct regions: The strip that extends along the ionization front from small radii ($<$ 0.5 AU) to high altitudes (40 AU) above the disk, and this emission is collisionally excited due to the presence of electrons, a proper temperature ($\sim$5000K), and an abundance of neutral oxygen. the other region extends above the surface of the disk and then flares up to about an angle of 60 from the pole; In both regions most of the emission comes from radii less than 3 AU. As can be seen there is an extended region beyond the stripe which includes some emission of the \forbid{Oi}{6300} line. In this region, the temperature is still appropriate for the thermal excitation of the O I line, the gas is mostly neutral, and the electron abundance (relative to H) is a few percent as can be seen from Fig. \ref{fig:Cloudy1}, so the emission is only a small fraction of the emission from the region along the strip.  \\

\begin{figure}[ht]
\centering
\frame{\includegraphics[scale = 0.525]{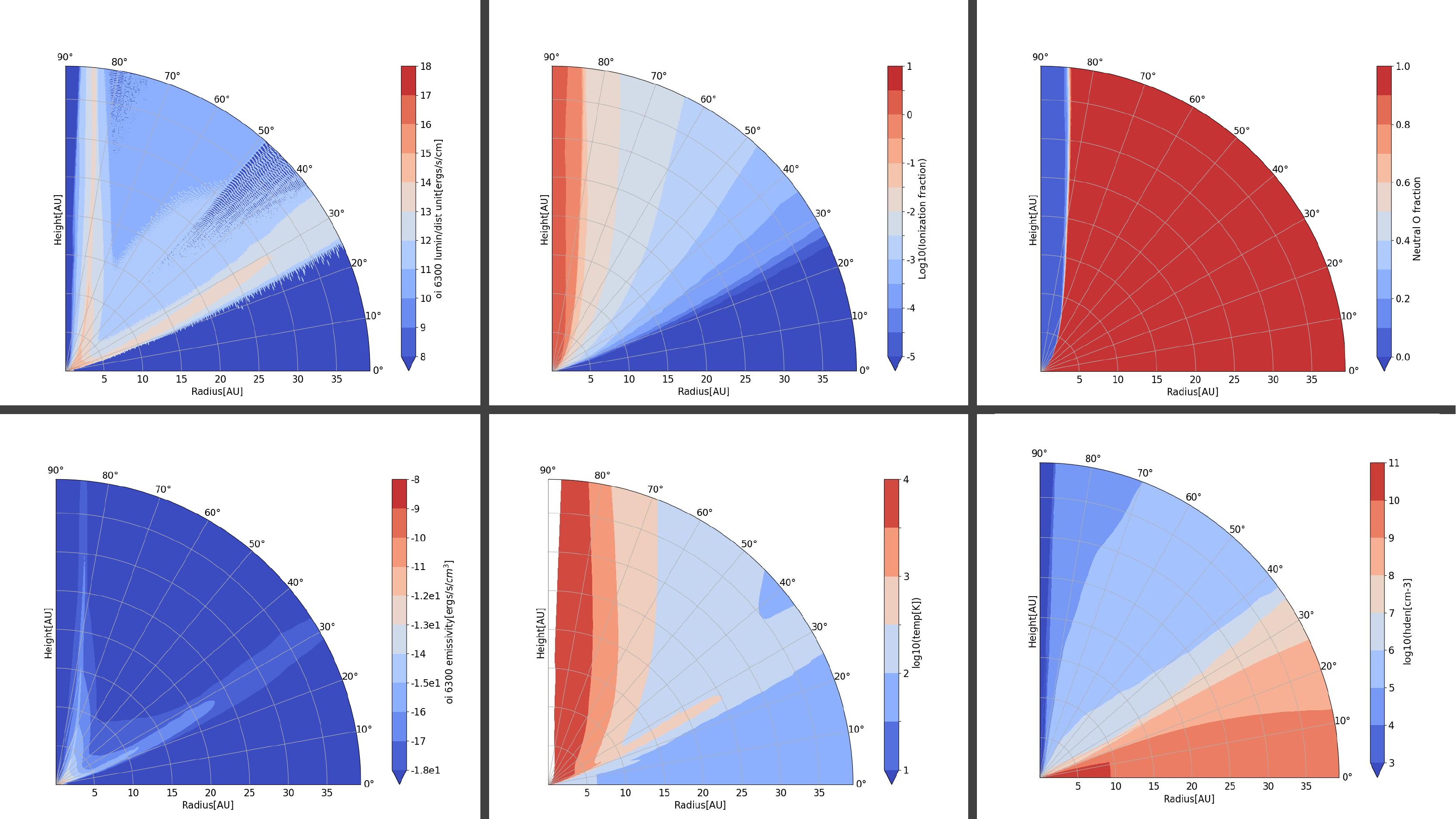}}
\caption{Polar plots of luminosity per distance and emissivity for \textsc{Oi} 6300 A along with the ionization fraction, neutral O fraction, temperature and density simulations with the 1 micron grains model using Owen et al model}
\label{fig:Cloudy1}
\end{figure}

\begin{figure}[ht]
\centering
\frame{\includegraphics[scale = 0.525]{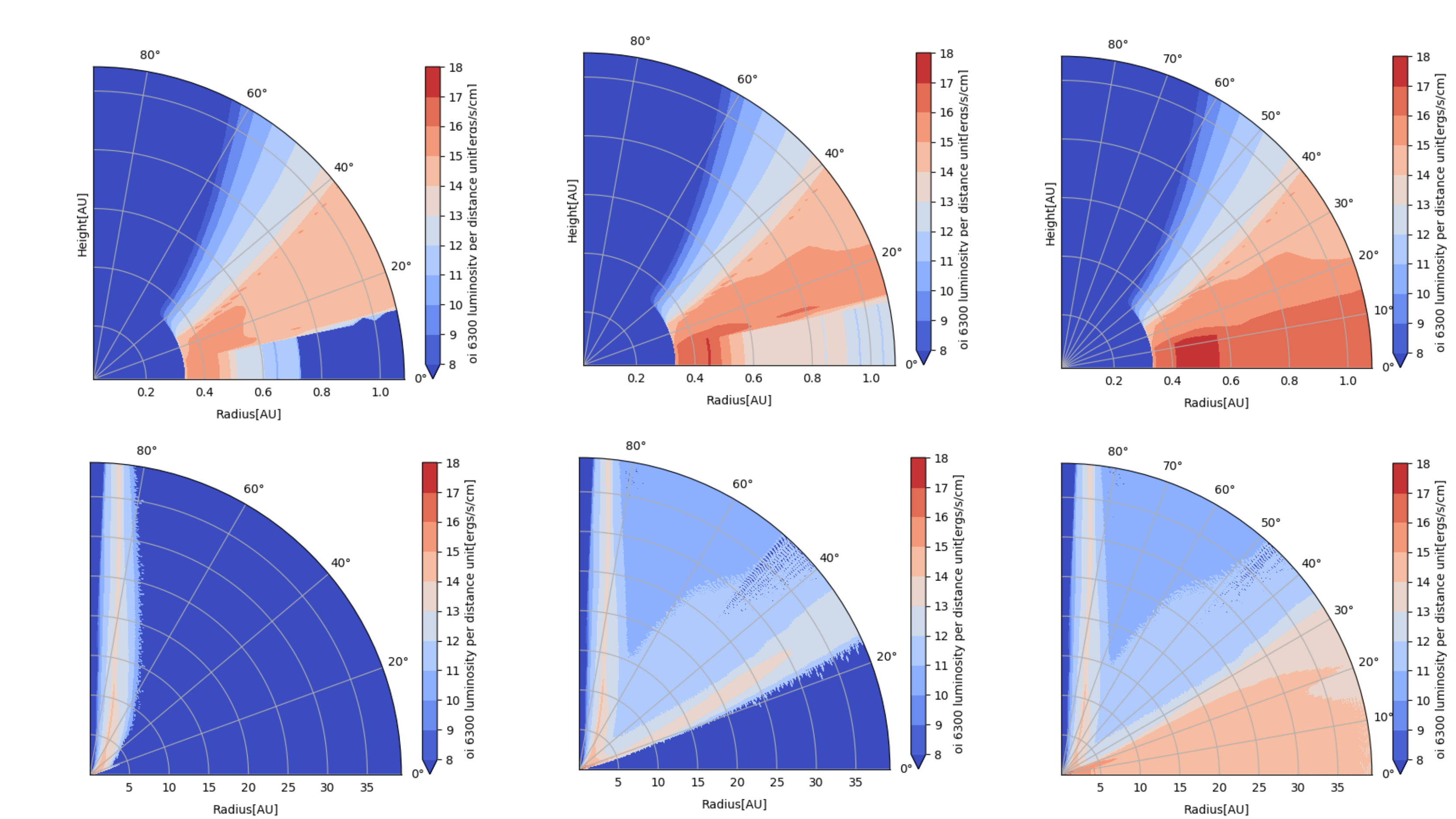}}
\caption{Polar plots for \forbid{Oi}{6300} luminosity using EO16's hydrodynamical model, zoomed into $1\au$, for three cases: The 1-$\micron$-grain model without pumping (left column); the corresponding model with pumping (middle column); and the grain-free model (right column). The luminosity per unit radial distance was produced by differencing the cumulative luminosity output by Cloudy, leading to some numerical noise.}
\label{fig:Cloudy2}
\end{figure}

We turn off the optical pumping for the fiducial model to find that the emission along the ionization front persists, but the emission from the region above the surface of the disk disappears(Fig. \ref{fig:Cloudy2}; turning off optical pumping is a function in Cloudy which disables optical pumping from the statistical equilibrium equations' solution. Moreover, the luminosities of the \textsc{Oi} 6300 and 5577 A lines drop by 80$\%$ and $90\%$, respectively as can be seen in table \ref{tab:lumin}. The ratio of the lines doubles to resemble the models that did not match the observed luminosity previously. The exception to the other cases is the model that uses no grains at all in the calculation. In that model we are able to produce the observed luminosity, and the majority of the emission, in addition to the above mentioned regions, comes from the disk mid-plane which is due to the complete lack of absorption of the FUV by the dust grains that would be concentrated in the disk mid-plane otherwise. \\

To verify the effect of the UV on the radiative transfer solutions, we reduce the luminosity of the two FUV sources (stellar and accretion) by an order of magnitude, while maintaining the X-ray luminosity. We notice that the ionization fraction does not change much throughout the wind although the ionization front recedes inwards towards the star due to the reduced ionizing radiation. The luminosity of the \textsc{Oi} 6300 line is reduced by 87$\%$ which tells us that the FUV has the dominant effect on the line emission as reported by EO16 and other authors. \\

In addition to heating and cooling interactions with the gas, dust grains absorb FUV radiation and re-emits it in the infrared as is evident by the correlation between infrared and FUV emission from YSOs \citep{Rigliaco+2013}. So, we traced the different types of radiation (IR, FUV, EUV, and Xray) as a function of distance from the star for certain directions in 1D simulations to understand the effect of dust on them. In all dust models mentioned above, the EUV radiation suffers from a significant drop at the ionization front (where all of the radiation is absorbed to ionizing the hydrogen and oxygen atoms) and the \textsc{Oi} 6300 emission peaks at that same point as discussed above. In the models where optical pumping did not have a significant effect on the \textsc{Oi} line emission, the FUV was quickly ($<$ 5 AU) converted into IR due to absorption and thermal emission by dust grains, and the \textsc{Oi} 6300 A emission drops significantly beyond the peak that coincides with the ionization front. On the contrary, in the models where optical pumping played a key role in the \textsc{Oi} line emission, the FUV was radially constant in those regions with significant [O I] emission, and had a higher luminosity than the IR. To confirm this result we had the code output the dominant processes for the various predicted lines, and it reported that all the neutral oxygen lines between 930 - 1350 A were predominantly populated through optical pumping by the FUV radiation (i.e. those lines are optically thick). The effect of this absorption is changing the population of the all the excited levels of neutral oxygen which in turn will change the emissivities of the \textsc{Oi} 6300 and 5577 A probably in different proportions.  \\

\begin{figure}[ht]
\centering
\frame{\includegraphics[scale = 0.525]{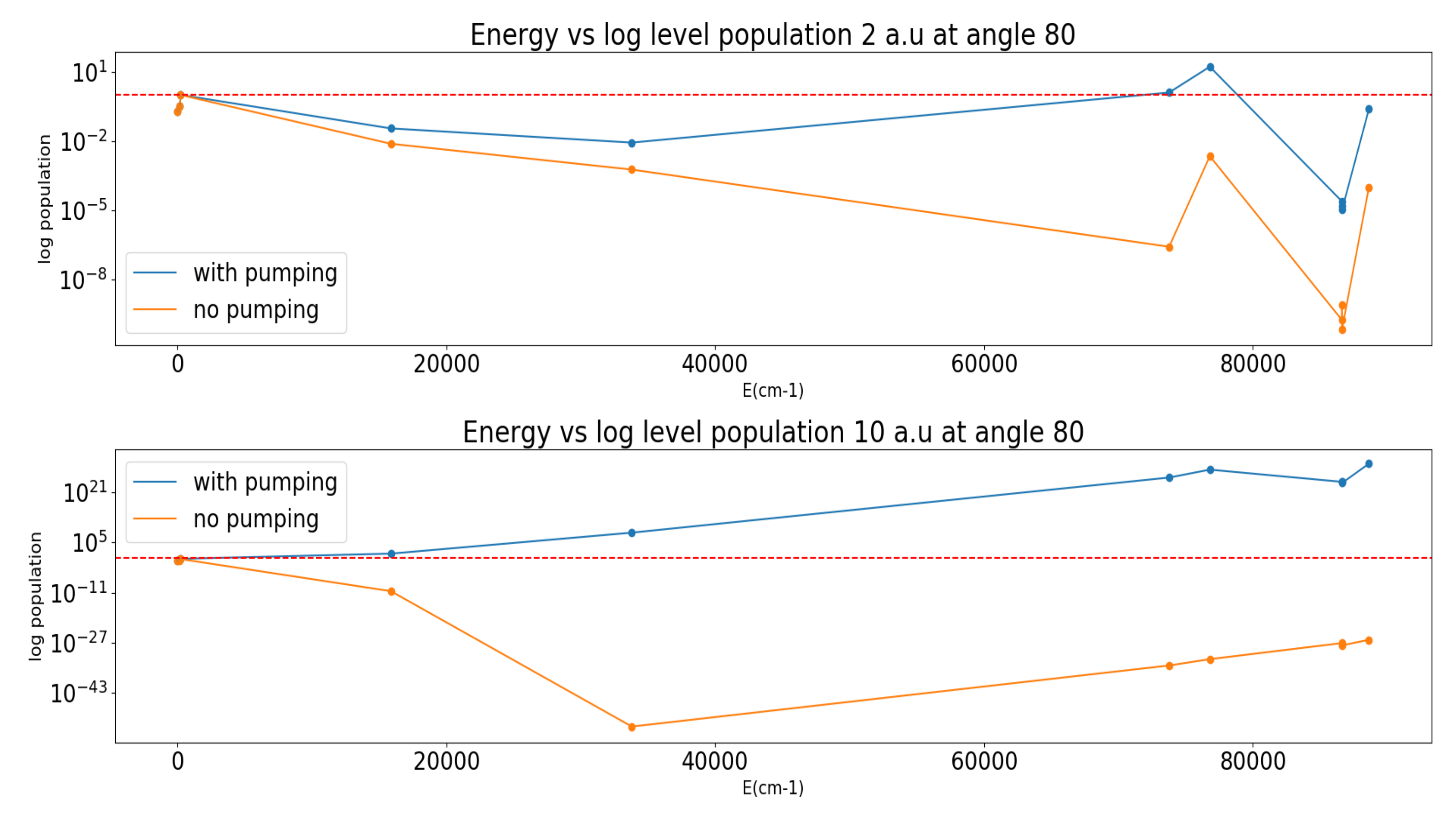}}
\caption{Level populations (divided by degeneracies and normalized for their LTE value) vs. energy above the ground for regions near the disk surface at (\emph{top}) $r=2\au$, $\nH=1.9\times10^7\unit{cm^{-3}}$, $T=3.9\times10^3\unit{K}$, $FUV_{E-density}=1.75\times10^{8}\unit{ergs/cm^2/s}$; and (\emph{bottom}) $10\au$, $8.7\times10^6\unit{cm^{-3}}$, $380\unit{K}$, $7.00\times10^{6}\unit{ergs/cm^2/s}$. The second \& third levels from the left are the upper levels of the \forbid{Oi}{6300} and 5575\AA{} lines.}
\label{fig:Cloudy3}
\end{figure}

To further demonstrate the effect of optical pumping on higher excited states of neutral oxygen, we show in Fig. \ref{fig:Cloudy3} the populations of the O atom's excited states as a function of their energy for three cases of the base model and using 1 micron grains for the dust: one where the excited states are forced to be populated according to their LTE values regardless of the physical conditions in their respective cells, another where all the atomic processes that control the populations of excited states, including optical pumping, and the case where we switch off the optical pumping and allow for all the other atomic processes. We normalized the non-LTE plots to the LTE case where the latter is represented by a horizontal line at unity and the latter by blue and orange curves. The behavior of the level populations are the same with and without pumping; except that the there is a general increase in the overall populations if you include pumping. The effect of FUV pumping on the forbidden line luminosities has been observed before \citep{Dupree2016} and will be discussed later. It is clear that the level populations in both models follow the collisional-radiative regime (which could obtain values higher than LTE case as seen in Fig. \ref{fig:Cloudy3}) and is hard to predict analytically \citep{Bar-Shalom1997,VanSijde1984}. These high excited levels are connected to each other through forbidden (change in J more than 1), semi-forbidden (change in spin), or allowed transitions. In the collisional-radiative regime where there is a complicated network of coupled transitions, the competition between collisions and radiative decay is subtle, and their solutions result in populations higher or lower than LTE. We note here that collisional transitions generally follow the electric-dipole selection rules that govern radiative transitions \citep{Khare2001}. \\

EO16 argue that much of the \forbid{Oi}{6300} emission comes from regions where collisions with atomic hydrogen rather than electrons dominate the excitation of the upper state. Such regions would have to have ionization fractions $<10^{-3}$ to balance the much higher ($>10^3$) collisional rate coefficients of electrons \citep{Simon+2016}. As can be seen from Fig. \ref{fig:Cloudy1}, the emission comes from regions where the ionization fraction is significantly higher than $10^{-3}$; EO16 report emission from similar spatial regions that has similar physical conditions of temperature, density, and ionization fraction. In addition, the ionization thresholds and the photoionization cross sections of neutral oxygen and hydrogen are similar, so we expect the gas to have insufficient neutral hydrogen in the emitting regions. EO16 and Cloudy use collision data with neutral hydrogen from the literature for the O I 6300 line \citep{Krems2006,Launay1977}, respectively. In addition, Cloudy includes an approximation for neutral H collision rate for the O I 5577 line \citep{Kiselman2000}, while EO16 don't. EO16 switched off the neutral H collisions for the \forbid{Oi}{6300} line, and they found that the line luminosity dropped by a factor of 3 which results in a drop in the ratio of O I lines by the same factor to match the observed ratio. We found that the effect of switching off collisions with neutral hydrogen had a negligible effect on the total luminosity of the \textsc{Oi} 6300 and 5577 lines; the luminosity only dropped by 2.5$\%$ and $<1\%$, respectively. \\


\begin{table}[ht]
\begin{center}
\begin{tabular}{|c|c|c|}
\hline
 Results  & Owen's model & Wang's model  \\
\hline
\forbid{Oi}{6300} luminosity ($L_{sol}$) & 1.2$\times10^{-5}$ & 8.99 $\times10^{-6}$ \\
\hline
\forbid{Oi}{6300} luminosity no pump ($L_{sol}$) & 1.8$\times10^{-6}$ & 5.84$\times10^{-6}$ \\
\hline
O I 5577 A luminosity ($L_{sol}$) & 1.0$\times10^{-6}$ & 3.37$\times10^{-6}$ \\
\hline
O I 5577 A luminosity no pump($L_{sol}$) & 3.7$\times10^{-8}$ & 2.99$\times10^{-6}$ \\
\hline
O I 6300/5577 ratio & 11.9 & 2.7 \\
\hline
O I 6300/5577 ratio no pump & 49.0 & 2.0 \\
\hline
\end{tabular}
\end{center}
\caption{Results from radiative transfer using Owen et al and Wang et al hydro-dynamical models using 1 micron grains. \\} 
\label{tab:lumin}
\end{table}

\subsection{Magnetothermal Wind}

To test the sensitivity of optical pumping to the density and temperature structure of the disk and wind, we have made a similar analysis of magnetothermal wind models like those of \citet{Wang2019} but extended to smaller radii in an attempt to confirm the role of optical pumping on the \textsc{Oi} forbidden lines. The radiation source was represented with five blackbody sources with various effective temperatures. We used a blackbody source with $T_{\rm eff}$ = 3900 K, M = 1.0 $M_{\rm sol}$ as the stellar source. Another source with $T_{\rm eff}$ = 9000 K was used to simulate the FUV radiation with a luminosity equal to a fraction of the stellar luminosity. Finally we included three more blackbody sources to mimic the X-ray and EUV radiation. One with $T_{\rm eff}$ = $1.85 x 10^5$ K, another source with  $T_{\rm eff}$ = $4.0 \times 10^5$ and another with $T_{\rm eff}$ = $3.4 \times 10^7$ K with  L= $10^{-1.6} \times L_{\rm bol}$ for EUV and L= $10^{-3} \times L_{\rm bol}$ for Xray. We use the same elemental abundances as reported by \citet{Wang2019} and we include dust and molecules in the model. The dust size and abundances are the same as the model that matched the observed luminosity of the \textsc{Oi} forbidden lines; namely, we used the 1 micron grains with an abundance of $\sim 1\%$. The gas starts at the surface of the star at r = $1.74 \times 10^{11}$ cm and ends at r = 4 AU. The density profile that was fed to these Cloudy models was extracted from the results of the magneto-hydrodynamic models computed following the work of \citet{Wang2019}. We conducted further simulations to investigate the innermost regions. The schemes of these simulations largely inherited \citet{Wang2019}, with a few additions to account for the processes taking place in the innermost disk. The updated thermochemical network includes K and K$^+$ (potassium and the positive ion) and their related reactions, especially collisional ionization that are believed to dominate the ionization of inner disk regions. The simulation has $240\times 144$ resolution for the 2.5D axisymmetric spherical polar grid covering grid $r\in[0.2~\au, 4~\au]$ and $\theta\in [0.06,\pi/2]$ respectively. The radial zones are spaced logarithmically, and the latitudinal zones are spaced in such a way that there are $\sim 10$ zones per scale height near the equitorial plane. Other numerical and physical conditions, including boundary and initial conditions of the disk and the fields, stellar properties, and high energy radiation luminosities, are setup in the the same way as \citet{Wang2019}. \\

\begin{figure}[ht]
\centering
\frame{\includegraphics[scale = 0.525]{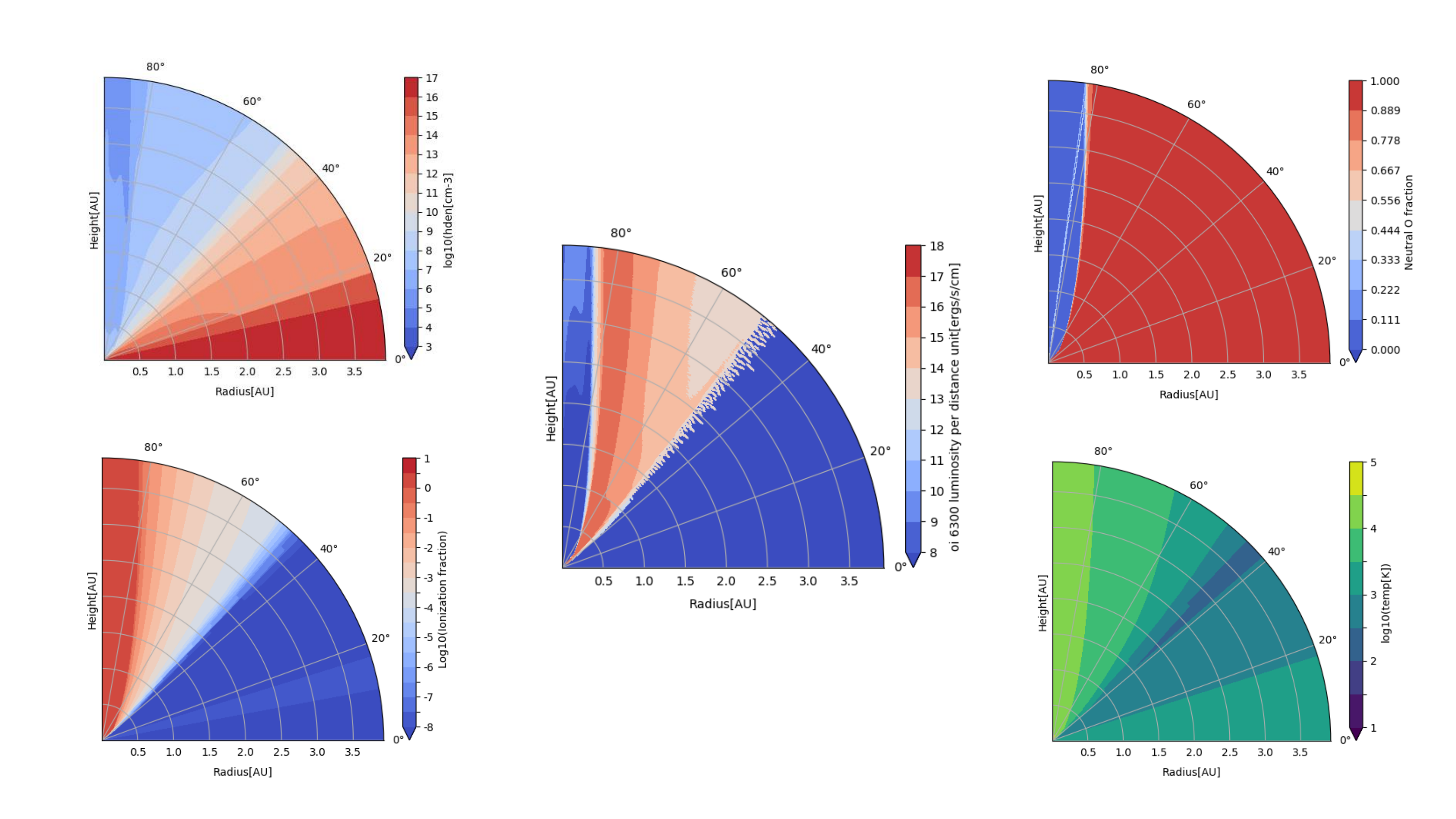}}
\caption{Polar plots of luminosity per distance for \textsc{Oi} 6300 A along with the ionization fraction, neutral O fraction, temperature and density in simulations with the 1 micron grains model using new MHD-wind models by Wang, following the methods of \cite{WBG2019} but focused on $r\le 5\au$.}
\label{fig:Cloudy1}
\end{figure}

We adopt the simulation results to study the effect of FUV pumping on the \textsc{Oi} 6300 and 5577 A lines: The first case that includes the above parameters; the second case where we switch off optical pumping for the base model. The results of these simulations are reported in table \ref{tab:lumin}. As can be seen the FUV pumping has a significant effect on the line luminosities, but in this case it affects the \textsc{Oi} 6300 A emission more. We note that the emission of the \textsc{Oi} 6300 and 5577 A lines are more prominent in the wind region and about 30 degrees angle from the pole.  

\section{Discussion} \label{sec:diss}

As can be seen from the results reported in the previous section, it is evident that the \forbid{Oi}{6300}:\forbid{Oi}{5577} emission is significantly affected by thermal excitation, but we showed for the simulations based on EO16's model that collisions are not the dominant excitation mechanism. Moreover, the conditions for H collisions to dominate over electron collisions are not available in the emitting regions reported by us and EO16 which makes it unlikely to have a significant effect on the line emission. \\

In both hydrodynamical models that we conducted radiation transfer for, we found that the FUV pumping affects the \forbid{Oi}{6300}:\forbid{Oi}{5577} emission appreciably. Using Owen et al's model, we found that the FUV pumping had the dominant effect on the line emission (80\% - 90\%) and that the thermal excitation was only a small fraction of that. In contrast, Wang et al's model shows a dominant effect for thermal excitation in the \forbid{Oi}{5577} $(\sim 88\%)$ and a stronger effect on the \forbid{Oi}{6300} $(\sim 65\%)$ than FUV pumping; but still FUV pumping played a key role in the line emission. This is probably because the density profile in Wang et al model was much higher than EO16, and the number of collisions is directly proportional to the density. \\

Furthermore, we noticed that the thermal emission in both cases came from similar regions; the strip that follows the ionization front. But the FUV pumping illuminated different regions of the disk in the two models. In the case of Owen et al's model, the FUV pumping originated from the surface of the disk while in Wang et al's model it came from the wind region. The physical conditions for the emitting regions in both models are very similar; they had a H density of $\sim 10^8 - 10^9 cm^{-3}$ and the temperature varied between 1000's to 100's K while the peak emission came from regions with temperature of $\sim$ 5000K. \\

When the level populations are in the collisional radiative regime (as can be seen from Fig. \ref{fig:Cloudy3}) the excited level populations can be obtained by solving the complex statistical equilibrium equations. The inclusion of FUV pumping affects the population of levels with a triplet term by pumping electrons from the ground which in turn increases the populations in the triplet sequence and changes the inter-system collisional rates. UV surveys document the dominating strength of the O I resonance lines in cool luminous stars \citep{Ayres1995} and require the inclusion of photoexcitation and photoionization from hydrogen radiation in the calculation of oxygen level populations and ionization. This effect was also observed by \cite{Dupree2016}. They showed that the O I forbidden lines (which were thought to be analyzed correctly with LTE \citep{Kiselman2000}) are changed drastically when the  Bowen fluorescence mechanism (absorption of Ly$\beta$ by atomic oxygen). \\

Electron collisions and FUV optical pumping are not the only processes that may excite the forbidden oxygen lines.  As noted in SI, \cite{Gorti+2011} and \cite{Rigliaco+2013} have suggested that photodissociation may be important and would naturally produce the line ratio $6300:5700\approx$ 7:1 as observed. \cite{Fang+2018} have argued against this on the grounds that the \forbid{Sii}{4068} line shows similar velocity structure to that of the [\textsc{Oi}] lines and is collisionally excited under similar conditions. \cite{Gorti+2011} made this suggestion in the context of their analysis of TW Hya, which shows a negligible blueshift: in this system, the oxygen lines presumably come from the disk. If there are other systems in which OH dissociation dominates the [\textsc{Oi}] luminosity and occurs mainly in the wind, rather than the disk, comparison with likely wind mass-loss rates shows that OH would have to re-form many ($\sim 100$ times) in the wind.  We have made simple estimates that indicate that the required reformation rates are plausible, but we have not made a systematic study of the contribution of OH dissociation to the [\textsc{Oi}] lines in either of the wind models discussed in this paper.

\section{Conclusion} \label{sec:con}
We have presented the results of radiation transport simulations, using Cloudy, on protoplantary disk environment using two distinct types of hydrodynamical model; an X-ray driven photoevaporative wind model of Owen and collaborators, and a magnetothermal model by Wang et al. We found that FUV pumping had a dominant effect on the \forbid{Oi}{6300} and \forbid{Oi}{5577} lines in the photoevaporative wind model, and weaker, yet significant, effect on the higher density magnetothermal wind model. The FUV pumping causes a general increase in the level populations of the oxygen atom, and that increase is dependant on the local physical conditions of the emitting region. The emission region of thermal excitation of the forbidden lines was consistent with previous calculations where the emission comes from the gas that follows the ionization front and the conditions for collisional excitation of the lines are met. On the contrary, the emission region of the FUV pumped \forbid{Oi}{6300} and \forbid{Oi}{5577} lines differed from one model to the other depending on the density and temperature profiles of the model. As a result, we expect that the consideration of FUV pumping in the radiation transport simulations will significantly affect the strengths and shapes of the \forbid{Oi}{6300} and \forbid{Oi}{5577} lines. It is not clear yet whether these lines can be used as tracers of disk winds in protostellar disk environments and to distinguish between the different wind models which will be left for future work.

\acknowledgments
We thank James Owen and Barbara Ercolano for sharing unpublished details of their work, Edward Jenkins and Bruce Draine for advice on atomic processes, Gary Ferland for help with Cloudy.  This work was supported by NASA grant 17-ATP17-0094.

\bibliographystyle{aasjournal}



\end{document}